\documentclass[fleqn]{2017SCGE}
\usepackage{graphicx}
\usepackage{bm}
\usepackage{upgreek}
\usepackage{lineno}
\usepackage{hyperref}

\begin{document}

\title{Stochastic spin-orbit-torque device as the STDP synapse \\for spiking neural networks}{Stochastic spin-orbit-torque device as the STDP synapse for spiking neural networks}

\author{Haotian Li}{}
\author{Liyuan Li}{}
\author{Kaiyuan Zhou}{}
\author{Chunjie Yan}{}
\author{\\Zhenyu Gao}{}
\author{Zishuang Li}{}
\author{Ronghua Liu}{{rhliu@nju.edu.cn}}

\AuthorMark{Li H T}

\AuthorCitation{Li H T, Li L Y, Zhou K Y, et al}

\address{National Laboratory of Solid State Microstructures, Jiangsu Provincial Key Laboratory for Nanotechnology, \\and School of Physics, Nanjing University, Nanjing 210093, China}

\abstract{Neuromorphic hardware, as a non-Von Neumann architecture, has better energy efficiency and parallelism than the conventional computer. Here, with the numerical modeling spin-orbit torque (SOT) device using current-induced SOT and Joule heating effects, we acquire its magnetization stochastic switching probability as a function of the interval time of input current pulses and use it to mimic the spike-timing-dependent plasticity learning behavior like actual brain working. We further demonstrate that the artificial spiking neural network (SNN) built by this SOT device can perform unsupervised handwritten digit recognition with an accuracy of 80$\%$ and logic operation learning. Our work provides a new clue to achieving SNN-based neuromorphic hardware using high-energy efficiency and nonvolatile spintronics nanodevices.}

\keywords{Spin-orbit torque, Neuromorphic hardware, Spiking neural network, Stochastic magnetization reversal}

\PACS{75.76.+j, 84.35.+i, 75.78.Jp, 85.70.Ay}

\maketitle

\begin{multicols}{2}
\section{Introduction}\label{section1}
For the last two decades, artificial intelligence (AI) techniques have been exerting extraordinary influence over the way we live\cite{AI,AI3}. Nowadays, the most widely used algorithm of AI is the convolutional neural network~\cite{cnn}, but the tremendous amount of computation is hard to be satisfied with the traditional computer under Von Neumann architecture. Recently, inspired by the human brain, the spiking neural networks (SNNs) using neuromorphic hardware have been proposed for AI applications (e.g., efficient signal processing and speech recognition), and show significantly lower energy consumption and higher data process efficiency than comparable classical artificial neural networks (ANNs)~\cite{nmhardware,nmhardware2,MAANN}.
\Authorfootnote
SNNs are a special class of ANNs, where the neuronal units communicate using discrete spike sequences analogous to a biological neuron~\cite{SNNbio}. In the SNNs, the inputs to a spiking neuron are discrete spikes through synapses. The spikes can change the membrane voltage of postsynaptic neurons with a dynamic synaptic weight. The synaptic weight is the strength of a connection between two neurons, and it can be modulated pre- or postsynaptically. After a series of discrete spikes inputs, the spiking neuron will produce an output spike if the membrane voltage reaches a certain firing threshold. Thus the learning process corresponds to the synaptic weight adaptation during training to fit the desired behavior by temporal order and interval time between the spikes of presynaptic and postsynaptic neurons, also called spike-timing-dependent plasticity (STDP)~\cite{STDP}.

In recent years, some reports suggest that the spintronic nanodevices with short-term memory and subnanosecond scale nonlinear magnetodynamics are the promising candidates to mimic neurons and synapses in SNN because of their low power consumption, non-volatility, and stochastic behavior~\cite{SNNmtj4,jiang,STreview,am}. In spintronics, SOT devices, without requiring a large working current passing through the actual layer of the device, can achieve high endurance and ultrafast write and read operations~\cite{switching,switching2,SOTintro,SOTintro2}, which are expected to have better performance than the conventional spin-transfer torque devices~\cite{SOTsynapse,SOTsynapse2,SOTsynapse3}.

In this work, we report that a SOT device implements an artificial synapse based on current-dependent magnetization switching probability. We find that the temporal correlations between the spikes of presynaptic and postsynaptic neurons, which are the key to STDP, can be simulated by a pair of former pulse and latter pulse with opposite polarity and the adjustable interval time via the current-driven Joule heating effect in a SOT device. The magnetization switching probability as a function of the input current pulses can be used to mimic the STDP learning behavior. Furthermore, we build two physical SNNs using SOT devices as artificial synapses and achieve unsupervised handwritten digit recognition and logic operation learning tasks.

\section{Experiments and simulations}\label{sec:2}
\subsection{Experiments of SOT-induced magnetization switching}
\begin{figure}[H]
\centering
\includegraphics[scale=1.0]{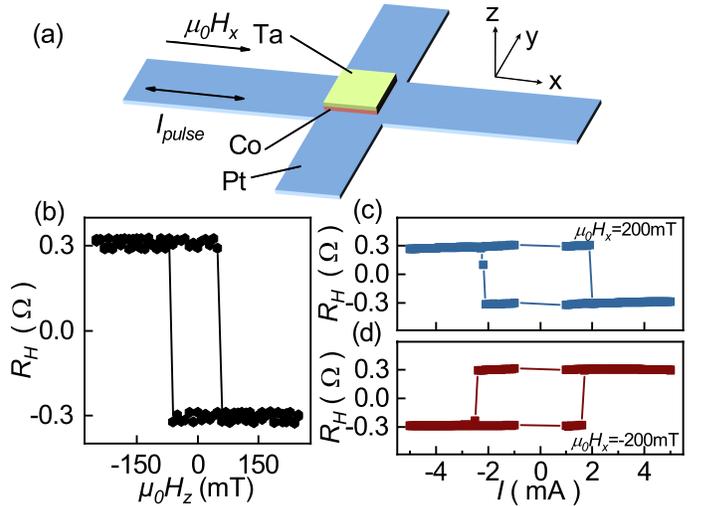}
\caption{(a) Schematic of the structure of the SOT Hall bar device and the experimental setup. (b) Hall resistance $R_{H}$ \emph{vs.} out-of-plane magnetic field $\mu_{0}H_{z}$ loop with a small current $I$ = 1 mA. (c)-(d) Current-induced magnetization switching of SOT device under an in-plane magnetic field parallel to the current flow direction $\mu_{0}H_{x}$ of (c) +200 mT and (d) -200 mT.}\label{fig1}
\end{figure}

Figure~\ref{fig1}(a) shows the schematic of a SOT Hall bar device and the definition of the coordinate. The multilayer stack is deposited on an oxidized Si wafer by dc magnetron sputtering with the following sequence: substrate/Ta(2)/Pt(5)/Co(0.8)/Ta(2), where the number in parentheses is the thickness in nm. The Ta(2)/Pt(5) bottom layers are first patterned into 3 $\mu$m $\times$ 10 $\mu$m cross Hall bar, and then a 2 $\mu$m $\times$ 2 $\mu$m square Co(0.8)/Ta(2) bilayer is fabricated at the center of the cross Hall bar by combining electron beam lithography and sputtering. As shown in Fig.~\ref{fig1}(b), the square Hall resistance hysteresis loop with the out-of-plane magnetic field reveals good perpendicular anisotropy of the magnetic film. Then we measure the current-induced magnetization switching of the device under an external magnetic field $H_{x}$ parallel to the current flow (the $x$-axis). Figures~\ref{fig1}(c) and ~\ref{fig1}(d) show examples of magnetization switching by applying $I$ under $\mu_{0}H_{x}$ = +200 mT and -200 mT. The magnetization begins to switch its direction when the applied dc current $I$ is over the critical value $I_{c}$ = 2 mA, and the switching polarity is determined by the direction of $\mu_{0}H_{x}$ and $I$~\cite{Ic,Ic2,MAic}.
\subsection{Simulations for STDP curves}
To demonstrate the SOT device with the capability of STDP for the application of SNN, we perform simulations with the experimental parameters to obtain the current-driven magnetization switching probability at different temperatures. Firstly, we can obtain current-induced SOT-driven magnetization switching properties of the Co layer by using open-source micromagnetic simulation software mumax$^{3}$ based on the generalized Landau-Lifshitz-Gilbert equation~\cite{mumax3,SOTLLG,ST}:

\begin{equation}\label{LLG}
  \dot{\textbf{m}} = -\gamma \textbf{m} \times (\textbf{H}_{eff} + a_{j}(\textbf{m} \times \boldsymbol{\sigma}) + b_{j} \boldsymbol{\sigma} ) + \alpha \textbf{m} \times \dot{\textbf{m}}
\end{equation}
where \textbf{m} is the reduced magnetization, $\gamma$ is the gyromagnetic ratio, $\alpha$ is the Gilbert damping constant, $\boldsymbol{\sigma}$ is the unit vectors of the spin polarization direction in the nonmagnetic heavy metal layers, $\textbf{H}_{eff}$ is the effective field including the external magnetic field $H$, the effective perpendicular anisotropy field $H_{k}$, and a fluctuating thermal field $H_{t}$~\cite{thermalfield2,thermalfield3}. $H_{t}$ is the Gaussian-distributed random fluctuation field with mean = 0 and the standard deviation = $\sqrt{2 \alpha k_{B} T / (\gamma M_{S} U \delta t)}$, where $k_{B}$ is Boltzmann constant, $T$ is the temperature of the ferromagnetic layer, $M_{s}$ is the saturation magnetization, $U$ is the volume of the ferromagnetic layer, and $\delta t$ is the integration time step~\cite{thermaltimestep}. In our Pt/Co/Ta-based Hall bar device, the spin polarization of the generated spin current in the bottom Pt and top Ta layers by the spin Hall effect is along the $y$-axis, perpendicular to the current flow direction. $a_{j}$ and $b_{j}$ correspond to the damping-like and field-like terms, respectively.

\begin{figure}[H]
	\centering
    \includegraphics[scale=0.95]{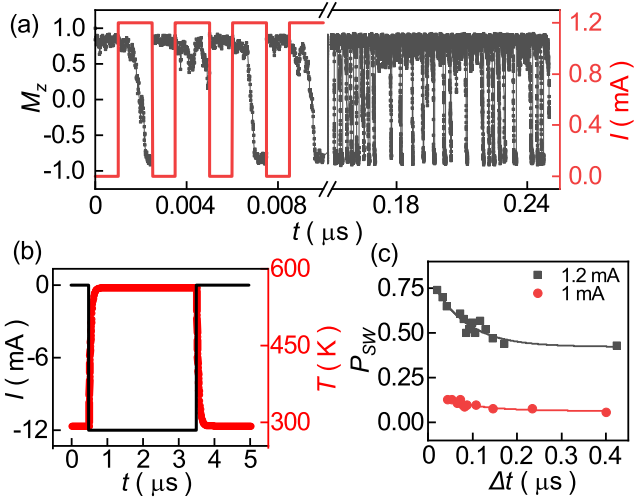}
	\caption{Micromagnetic and COMSOL simulation results. (a) The magnetization $M_{z}$ evolution of the SOT device with input current pulse. The duration width of the current pulse is 1.5 ns, and the period is 2.5 ns. To easily illustrate device switching probability, $M_{z}$ is reset to the initial state before each new repeat pulse. (b) The temperature of the device increases to 563 K and then decreases exponentially to 293 K in 0.25 $\mu$s when a square current pulse with $I_{reset}$ = -12 mA is applied to the device. (c) Magnetization switching probability $P_{sw}$ as the function of the interval time $\Delta t$, defined in the text, at $I_p$ = 1 mA and 1.2 mA. The solid lines are the exponential function fitting curves. }\label{fig2}
\end{figure}

The input current pulse with width $\tau$ = 1.5 ns, amplitude $I_{p}$ = 1.2 mA and 100 repeat cycles are used to determine the switching probability of the Co magnetization due to the spin current exerting spin torques. The spin current can be calculated by $I_{p}$ and spin Hall angle $\alpha_{H}$. The parameters used in the micromagnetic simulation are listed in Table \ref{tab:table1}. The initial direction of magnetization is defined parallel to the $z$-axis ($M_{z} = +1$), and the $M_{z}$ will be reset to the initial state before each new repeat pulse. Figure~\ref{fig2}(a) shows a representative $M_{z}$ \emph{vs.} time $t$ curve (symbols) under a square current pulse (red line) and the external in-plane magnetic field $H_{x}$ = 10 kA/m. One can see that the magnetization $M_{z}$ exhibits the probabilistic flip event at $I_{p}$ = 1.2 mA, slightly smaller than the critical current $I_{c}$, because the considered fluctuating thermal field $H_{t}$ causes the stochastic magnetization switching~\cite{thermal}. The switching probability can be described by $P_{sw} = 1 - {\rm exp}(-f_{0} \tau {\rm exp}(-E_{B}/k_{B}T))$, where $f_{0}$ is the attempt frequency, $\tau$ is the current pulse width, $T$ is the temperature of the magnetic system, and $E_{B}$ is the energy barrier~\cite{probability}.

Next, the temperature of the ferromagnetic Co layer as a function of the time after the input current pulse can be accurately simulated by the electromagnetic heat module of COMSOL Multiphysics$^{\circledR}$. The actual Ta(2)/Pt(5)/Co(0.8)/Ta(2) cross Hall bar-based SOT device is modeled. The substrate SiO$_{2}$ is used as the heat sink with a reference temperature $T_{ref}$ = 293.15 K. As injecting a current pulse into the SOT device, the Joule heat generated by the input current is localized in the metallic films due to the limited size of the cross Hall bar and interfacial thermal resistance between the bottom Ta and the heat sink SiO$_{2}$. As a result, the temperature of metallic films will dramatically increase and ultimately reach a saturation temperature after the thermal equilibrium process. For the inverse process, the temperature of metallic films will sharply drop to $T_{ref}$ when the current pulse ends. Based on the previous report~\cite{COMSOL1}, the gap heat conductance $h_{g}$ between the bottom Ta and the heat sink SiO$_{2}$ is adopted as 600 MW/(m$^{2}\cdot$K). All parameters used in the COMSOL calculation are also summarized in Table~\ref{tab:table1}. Figure~\ref{fig2}(b) shows the representative temperature evolution of the SOT device with the input square current pulse, well consistent with the previously reported results~\cite{COMSOL1}.

To illustrate the STDP of our Hall bar-based SOT device, we need two square current pulses with different amplitudes and polarities in one cycle. The preceding negative pulse $I_{reset}$ far higher than $I_{c}$ is used to reset the $M_{z}$ to the initial state with 100$\%$ switching probability and produces a significant Joule heat-induced temperature rise, while the following positive pulse $I_p$ with an amplitude below $I_{c}$ causes the stochastic switching and a tiny Joule heating. We define $\Delta t$ as the interval time between the negative initial pulse and positive stochastic switching pulse, in analogy to the interval time between the spike event of neurons. Therefore, the temperature variation of the Co layer caused by the current-induced Joule heating effect is a function of $\Delta t$.

\begin{table}[H]
\footnotesize
\begin{threeparttable}
\caption{\label{tab:table1} Simulations parameters.}
\doublerulesep 0.1pt \tabcolsep 7pt
\begin{tabular}{ccc}
\toprule
Parameters & Description & Default Values \\\hline
$M_{s}$    & Saturation magnetization (A/m) & $1\times 10^{6}$ \\
$a_{H}$    & Spin Hall angle & 0.07 \\
$\alpha$    & Gilbert damping factor & 0.03 \\
$H_{x}$  & External magnetic field (A/m) & $1\times 10^{4}$ \\
$A_{ex}$   & Exchange coefficient (A$\cdot$ m) & $1\times 10^{-11}$ \\
$H_{k}$    & Perpendicular anisotropy field (A/m) & $1\times 10^{5}$ \\
$T_{ref}$  & Reference temperature (K) & 293.15 \\
$h_{g}$    & Gap heat conductance (W/(m$^{2}$$\cdot$K))& $6\times 10^{8}$ \\
$C_{Pt}$    & Heat capacity of Pt (J/(kg$\cdot$K))& $500$ \\
$C_{Co}$    & Heat capacity of Co (J/(kg$\cdot$K))& $420$ \\
$C_{Ta}$    & Heat capacity of Ta (J/(kg$\cdot$K))& $140$ \\
$\rho_{Pt}$    & Resistivity of Pt ($\mu\Omega$ cm)& $30$ \\
$\rho_{Co}$    & Resistivity of Co ($\mu\Omega$ cm)& $40$ \\
$\rho_{Ta}$    & Resistivity of Ta ($\mu\Omega$ cm)& $200$ \\
\bottomrule
\end{tabular}
\end{threeparttable}
\end{table}

Now, we can obtain the switching probability as the function of $\Delta t$ by micromagnetic simulation with Eq.\ref{LLG}, including Gaussian-distributed random fluctuation thermal field $H_{t}$. For a specific $\Delta t$, the temperature of the Co layer corresponding to $H_{t}$ can be obtained by the COMSOL calculation above. The magnetization switching probability as a function of $T$ is further determined by micromagnetic simulation. Figure~\ref{fig2}(c) shows two representative curves of the switching probability as the function of $\Delta t$ at $I_p$ = 1 mA and 1.2 mA. One can see that the smaller $\Delta t$, corresponding to the shorter interval time between the positive stochastic switching pulse and the initial pulse, exhibits the higher magnetization switching probability $P_{SW}$ due to the larger Joule heat-induced temperature rise. These two curves can be well fitted by the exponential function $P_{sw} = 1 - {\rm exp}(-f_{0} \tau {\rm exp}(-E_{B}/k_{B}T))$~\cite{probability}. The fitting curves will be used as the STDP rules of SNN discussed below.

\section{Neuronal dynamics theory of SNN}
\subsection{Leaky integrate-and-fire model of neurons}
The SNNs, as a type of ANNs, consist of a couple of layers of spiking neurons interconnected by synapses with the STDP rule. The information transferred between different neurons is carried by electrical spikes($i.e.$ spiky pulses) through synapses. As illustrated in Fig.~\ref{fig3}(a) for a simplified neuronal system, the grey circle represents a neuron (e.g., postsynaptic neuron), which connects with three other neurons (e.g., presynaptic neurons) via synapses. The difference between the interior and exterior electric potentials of a neuron is called membrane potential $V$. It usually remains constant as the resting membrane potential $V_{reset}$ until the neuron is stimulated by spikes. As shown in Fig.~\ref{fig3}(b), the membrane potential $V$ as a function of time can be described by leaky-integrate-and-fire (LIF) model~\cite{LIF}:

\begin{equation}\label{LIF}
  \tau\frac{dV}{dt} = (V_{rest}-V)+g_{e}(V_{exc}-V)+g_{i}(V_{inh}-V)
\end{equation}
where  $V_{rest}$ is the resting membrane potential, $V_{exc}$ and $V_{inh}$ are the equilibrium potentials of excitatory and inhibitory synapses, $g_{e}$ and $g_{i}$ the conductances of excitatory and inhibitory synapses, and $\tau$ is a time constant, respectively. In the process of transferring information, when a neuron receives spikes from other neurons through synapses, its membrane potential jumps abruptly and then decreases exponentially with time [top half of Fig.~\ref{fig3}(b)], as described by Eq.\ref{LIF}. Especially, a neuron emits a spike(\emph{i.e.,} a neuron fires) when its membrane potential is higher than the threshold voltage $V_{threshold}$, $V$ is reset to $V_{reset}$, and then the neuron can not be stimulated over a refractory period.

\subsection{STDP rule of synapses}
\begin{figure}[H]
	\centering
	\includegraphics{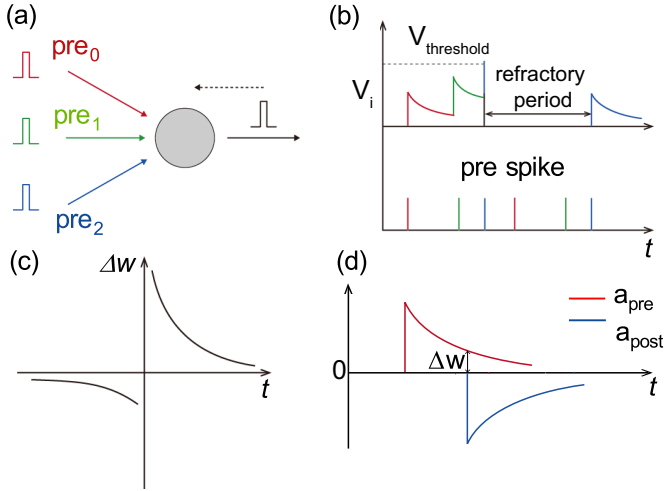}
	\caption{(a) Schematic representation of the neuron (grey circle) and three synapses(colored arrows) are stimulated by the input spike trains (pre$_{i}$ ($i$=0, 1, and 2)). (b) The temporal course of the membrane potential of a neuron in (a). The neuron receives the spike sequences from presynaptic neurons, increasing membrane potential. Once the membrane potential is over the threshold $V_{threshold}$, the neuron will emit a spike with a short refractory period. (c) Weight change curves for the standard STDP. The curve in the top right corner represents the LTP corresponding to the fitting curve of $I_{p}$ = 1.2 mA in Fig.~\ref{fig2}(d) without constant item, while the LTD curve in the bottom left corner corresponds to the fitting curve of $I_{p}$ = 1 mA in Fig.~\ref{fig2}(d). (d) The $a_{pre}$ and $a_{post}$  are the ``traces'' of presynaptic and postsynaptic activity. The $a_{pre}$ or $a_{post}$ increases abruptly when the presynaptic or postsynaptic neurons fire and then decrease slowly with time. At the firing point of the postsynaptic neurons, the weight change of the synapse is equal to $a_{post}$ - $a_{pre}$. }\label{fig3}
\end{figure}

In a neural network, a postsynaptic neuron connects with others through a number of synapses. The neuron will fire when its membrane potential reaches the threshold after received continuous spikes from other connected neurons through synapses, as shown in Fig.~\ref{fig3}(b). The effect of spikes on the postsynaptic neuron is determined by the strength of synapses, $i,e.$ weight. In other words, the generated variation of $V$ for the postsynaptic neurons by spikes from the presynaptic neurons is proportional to the connecting weights of synapses. The learning process is the synaptic weight adaptation following the discussed STDP rule below. If the firing time of presynaptic neurons($t_{pre}$) is before the firing time of postsynaptic neurons($t_{post}$), the weight of this synapse will increase, or else the weight will decrease. This firing time sequence expresses the causality of the neural network. In general, the value of the updated synaptic weight $\Delta w$ is a function of the relative timing between presynaptic and postsynaptic neurons firing~\cite{STDPrule,MAstdp}:

\begin{flalign}\label{STDP}
    {\triangle w = }
    \left\{
        \begin{array}{l}
            A^{+} e^{\frac{\Delta t}{\tau_{+}}}, \  \Delta t < 0 \\  A^{-} e^{\frac{\Delta t}{\tau_{-}}}, \  \Delta t \geq 0
        \end{array}
    \right.
        \end{flalign}
where $\Delta t = t_{post}-t_{pre}$, $A_{+} > 0$ and $A_{-} < 0$ are constant, and time constant $\tau _{+}, \ \tau_{-} > 0$. The top and bottom terms in Eq.\ref{STDP} are long-term potentiation(LTP) and long-term depression terms(LTD), which correspond to the top-right curve and bottom-left curve in Fig.~\ref{fig3}(c), respectively. To implement this STDP rule of synapses by using a physical device or material, we adopt the concept of ``trace'' to express the value of the synaptic weight $\Delta w$, the same as the previous works used the python package Brian 2~\cite{brian2}. As shown in Fig.~\ref{fig3}(d), the two traces $a_{pre}$ and $a_{post}$ correspond to presynaptic activity and postsynaptic activity, respectively. Suppose that a presynaptic neuron fires, $a_{pre}$ increases by $A_{+}$ abruptly and then declines exponentially with time. While $a_{post}$ will decrease by $|A_{-}|$ and fades exponentially with time when the postsynaptic neuron fires. As a result, the update rule of synaptic weight with time can be implemented by two physical ``trace'' curves $a_{pre}$ and $a_{post}$. In other words, magnetization switching probability $P_{SW}$ of SOT device is the function of the interval time $\Delta t$ in Fig.~\ref{fig2}(c).

Therefore, we can use the obtained switching probability \emph{vs.} $\Delta t$ curves of the SOT device as the STDP rule of the SNNs. As shown in Fig.~\ref{fig3}(c), for the standard STDP curve, the $\Delta t$ can be in analogy to the interval time of the fired neurons, and the weight change $\Delta W$ is reflected by the switching probability of the SOT device, which is in proportion to its electrical conductance. Below we use two tasks of the unsupervised handwritten digits recognition and logic operation learning as an example to test and verify the performance of the SNN built with the SOT device.

\section{Results and discussion}\label{sec:3}
\subsection{SOT devices as synapses for handwritten digital recognition}
\begin{figure}[H]
	\centering
	\includegraphics[scale=0.95]{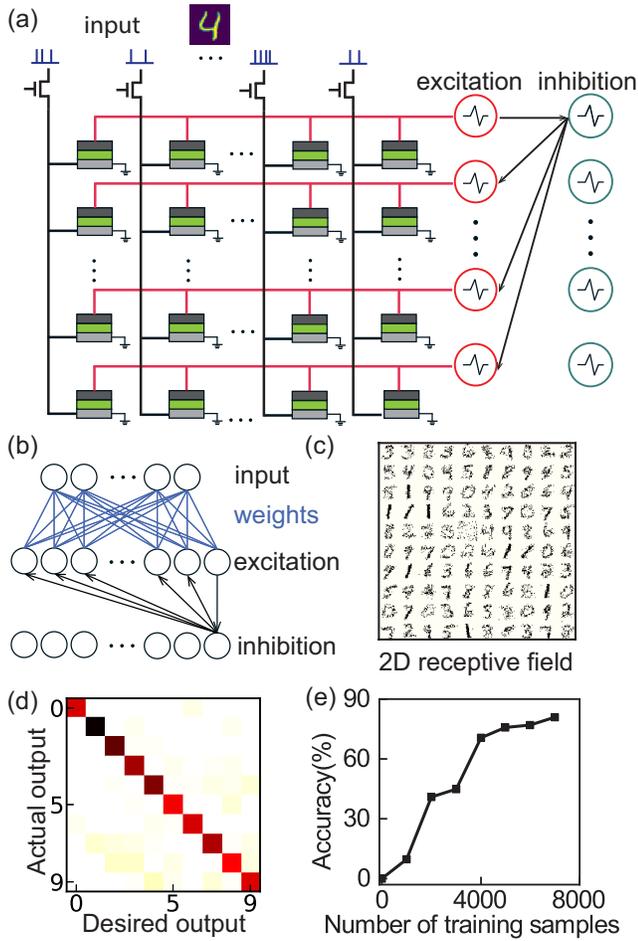}
	\caption{(a) Cross-bar array of SOT devices for pattern recognition based on SNN. The architecture contains three kinds of neurons and synapses implemented by SOT devices. The intensity values of the 28 $\times$ 28 pixels MNIST image (e.g., number 4) are converted to Poisson-spike with the firing rate proportional to the intensity of the corresponding pixel. Then the SOT devices, as the synapses, receive the input pulses signal through controlled transistors and transmit the weighted pulses to excitatory neurons. (b) The schematic of SNN for pattern recognition. (c) Table of handwritten digital numbers with 10 $\times$ 10 grids, where each grid displays a digital number with 28 $\times$ 28 pixels, as the learning result of synaptic weights $w$ between the neurons in the input and excitatory layer. The synaptic weight $w$ only can take the specific two values 0.01 (represented by white) and 1 (represented by black), which correspond to the two magnetization states of SOT devices. (d) Handwritten digit recognition accuracy rates with sampling 10,000 MNIST testset images. The $x$-axis is the desired digits and the $y$-axis is the predicted results. (e) The recognition accuracy rates as a function of the training sample counts.}\label{fig4}
\end{figure}

We build the physical SNN with SOT devices under the STDP rule to test its performance in the standard handwritten numbers recognition task using the Mixed National Institute of Standards and Technology (MNIST) dataset~\cite{SNNprogram,MNIST}. Figure~\ref{fig4}(a) shows the process flow diagram of the SNN for handwritten digits recognition, which consists of an input layer containing 28$\times$28 neurons corresponding to the 28$\times$28 pixels of a handwritten digit image, an excitatory layer (EL) of 100 neurons, and an inhibitory layer (IL) with corresponding 100 neurons providing lateral inhibition to 100 neurons in the excitatory layer. In the physical SNN, many previous reports mentioned that the neurons in SNN can be implemented by CMOS, memristors or other spintronic nanodevices~\cite{neuronsMemristor,neuronsCMOS,SNNmtj3,SNNmtj2}. The neurons of the input layer are connected to that of the EL with synapses based on SOT devices in an all-connection fashion. For an MNIST digit image, e.g., digital number ``4'', each neuron of the input layer encodes the pixel value of the image into the form of Poisson-distributed spike temporal sequences of 350 $\mu$s, and the firing rate of the neuron is proportional to the pixel value. Then the spike temporal sequences controlled by transistors pass through the STDP synapses and act on the EL. Figure~\ref{fig4}(b) is a sketch of the SNN architecture. In the training phase, since all synaptic weights have random initial values, the EL neurons will be fired with different rates for a particular image ``4'' as the input. The neuron with the highest firing rate is assigned to the output digital number ``4'' represented by the input image. Meanwhile, the neuron marked as ``4'' sends a spike to the corresponding neuron in the IL and fires it to inhibit all other 99 neurons in the EL and realizes the winner-take-all mechanism in the SNN.

In principle, higher accuracy needs more the EL neurons to carry the features corresponding to the different handwritten digit images representing the same digital number. However, considering the trade-off between the benefits of the recognition accuracy and the cost of computation, we adopt 100 EL neurons and corresponding 100 IL neurons for this test case. Figure~\ref{fig4}(c) shows the obtained training weights of all-connected synapses between the input layer with 28$\times$28 neurons and the 100 EL neurons, which are rearranged into 10$\times$10 grids and each grid contains 28$\times$28 weight values. These weights only can have two specific values of 0.01 and 1, which can be implemented using two Hall resistance states of SOT devices. The Hall resistance states are determined by the STDP rules based on switching probability. After the training process, all synaptic weights (corresponding to magnetization configurations of SOT device arrays) are determined and fixed. In the test phase, a new test image is fed into the SNN with the determined weights. The EL neurons with the same marked digital number have the highest average firing rate in the EL as the output result. Figure~\ref{fig4}(d) shows the comparison between the predicted results and the desired outputs for 10 digital numbers. One can see that the numbers ``7'' and ``9'' exhibit a relatively high error rate of recognition due to their high shape similarity in the form of random handwritten digit images. Figure~\ref{fig4}(e) shows that this SNN with 100 EL neurons can reach a better than 80\% accuracy in the unsupervised handwritten digit recognition task when the number of training samples is over 4000.
\begin{figure}[H]
	\centering
	\includegraphics{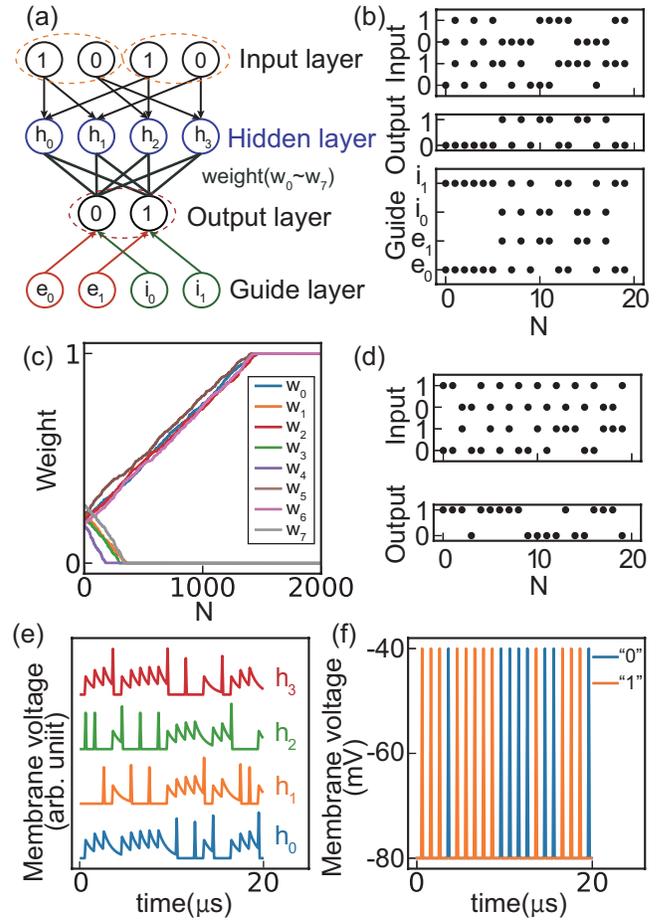}
	\caption{(a) SNN hierarchical architecture for logic operation learning. The input layer converts the logic variables to spikes for subsequent learning, and the guide layer contains excitatory and inhibitory neurons, which control the membrane voltage of output neurons in the training phase. (b) The used data for XOR operation in the training phase. (c) The updated process of the synapse weights $W_0 - W_7$ between the hidden and output layers with feeding 2000 training samples. All eight weights are determined after the training samples N $\geq$ 1500, indicating that the SNN completes its learning or training procedure. (d) 20 pairs of random logic numbers as inputs and the desired results of XOR operation in the testing phase. (e) The membrane voltage of all neurons in the hidden layer with input sequences. The neuron fires only when both fired input neurons are connected to it. (f) The membrane voltage of all output neurons in the test phase. When the output is 0, the corresponding neuron-0 will emit a spike as the result and vice versa.}\label{fig5}
\end{figure}
\subsection{SOT devices as synapses for logic operation learning}
Besides the handwritten digit recognition task above, we also choose logic operation learning to explore the additional function of our artificial SNN with SOT devices~\cite{logicsnn}. XOR is one of the sixteen possible binary operations of Boolean operations. Here, we choose the XOR learning as an example to illustrate how the SNN works. Other fifteen logic operations (e.g., NAND and NOR) are also able to be easily achieved in a similar way by our proposed physical SNN. Figure~\ref{fig5}(a) shows the SNN architecture for the binary logic operation learning task, which consists of an input layer containing 2$\times$2 neurons corresponding to two Boolean inputs and each with two different states ``True'' and ``False'', a hidden layer (HL) with four hidden neurons, and an output layer (OL) with two output neurons labeled ``1'' and ``0'' corresponding the binary results of a Boolean output. Different from the conventional binary encoding method in SNN where the neuron emits a spike usually represents the logic ``True'' and without spike indicates the logic ``False'', here, we define the specific neurons in the input and output layers as ``True'' or ``False'' neurons beforehand, as labeled using ``1'' and ``0'', respectively [Fig.~\ref{fig5}(a)]. For example, if the neuron labeled ``0'' (``1'') in the dotted oval emits a spike, the input is a logic ``False'' (``True''). It is the same for the output layer. The function of two excitatory neurons (``$e_0$'',``$e_1$'') and two inhibitory neurons (``$i_0$'', ``$i_1$'') in the guide layer (GL) is to train the synaptic weights between the HL and OL in the training phase. The firing time of the GL neurons ``$e_{0,1}$'' and ``$i_{0,1}$'' are controlled by an externally supervised signal; therefore, the STDP function can be realized by adjusting the spike temporal sequences generated from the HL and GL. Here we will illustrate this process by a specific example of both ``False'' (``0'') inputs and XOR ``False'' output, as shown in Fig.~\ref{fig5}(b). The XOR ``False'' output means that neuron-0 needs to be fired in the OL. Therefore, all weights of neuron-0 with the HL need to be enhanced and all weights from the HL to neuron-1 need to be reduced. To achieve this desired adjustment of weights, the neuron $e_{0}$ ($i_{1}$) is controlled by a supervise signal to emit a spike behind (before) the spike fired by the HL, corresponding to the STDP curve at $\Delta t>0$ ($\Delta t<0$)[Fig.~\ref{fig3}(d)]. We use $N$ to represent the number of input samples. Figure~\ref{fig5}(c) shows the evolution of the eight synaptic weights with increasing the training samples. One can see that all weights become stable (0.01 or 1) after $N \geq $1500. In the testing phase, as shown in Fig.~\ref{fig5}(d), we choose 20 pairs of random logic numbers as the input to test the SNN with the determined weights. Figures~\ref{fig5}(e) and~\ref{fig5}(f) show the membrane voltage of all neurons in the HL and OL as a function of input sequences. In the hidden layer, the neuron fires when both inputs are connected to it. The labeled logic variable of the fired neuron of the output layer is the testing result of the XOR operation, e.g., the given result is 0 if neuron-0 of the output layer fires. It is observed that the sequence of the fired neurons in the OL, shown in Fig.~\ref{fig5}(f), is well consistent with the desired output in Fig.~\ref{fig5}(d), proving the validity of the logic operation learning with our SOT-device-based SNN.
\section{Summary}\label{sec:4}
In summary, we propose an artificial synapse numerical model based on Pt/Co/Ta-based SOT device and use it to build the physical SNNs. In these physical SNNs, the adaption process of synaptic weight follows the STDP behavior of the SOT device, obtained by micromagnetic simulation of the current pulse and its interval time tunable magnetization stochastic switching probability of the SOT device due to current pulse-induced SOT and Joule heating effects. Our physical SNNs exhibit good performance in unsupervised handwritten digit recognition with over 80$\%$ accuracy and logic operation learning. Our work offers a new clue for spintronic hardware implementation of neuromorphic computing systems.

\Acknowledgements{This work was supported by the National Natural Science Foundation of China (Grant Nos. 12074178), and the Open Research Fund of Jiangsu Provincial Key Laboratory for Nanotechnology.}

\end{multicols}

\begin{thebibliography}{99}

\bibitem{AI}Y. LeCun, Y. Bengio,and G. Hinton, Nature \textbf{521}, 436 (2015).
\bibitem{AI3}D. Silver, A. Huang, C. J. Maddison, A. Guez, L. Sifre, G. van den Driessche, J. Schrittwieser, I. Antonoglou, V. Panneershelvam, M. Lanctot, S. Dieleman, D. Grewe, J. Nham, N. Kalchbrenner, I. Sutskever, T. Lillicrap, M. Leach, K. Kavukcuoglu, T. Graepel, and D. Hassabis, Nature \textbf{529}, 484 (2016).
\bibitem{cnn}R. Yamashita, M. Nishio, R. K. G. Do, and K. Togashi, Insights into Imaging \textbf{9}, 611 (2018).
\bibitem{nmhardware}C. D. Schuman, S. R. Kulkarni, M. Parsa, J. P. Mitchell, P. Date, and B. Kay, Nat. Comput. Sci. \textbf{2}, 10 (2022).
\bibitem{nmhardware2}K. Roy, A. Jaiswal, and P. Panda, Nature \textbf{575}, 607 (2019).
\bibitem{MAANN}Y. Zhang, Q. Zheng, X. Zhu, Z. Yuan, and K. Xia, Sci. China Phys. Mech. Astron. \textbf{63}, 277531 (2020).
\bibitem{SNNbio}A. Taherkhani, A. Belatreche, Y. Li, G. Cosma, L. P. Maguire, and T. M. McGinnity, Neural Netw. \textbf{122}, 253 (2020).
\bibitem{STDP}H. Markram, Science \textbf{275}, 213 (1997).
\bibitem{SNNmtj4}A. Mizrahi, T. Hirtzlin, A. Fukushima, H. Kubota, S. Yuasa, J. Grollier, and D. Querlioz, Nat. Commun. \textbf{9}, 1533 (2018).
\bibitem{jiang}W. Jiang, L. Chen, K. Zhou, L. Li, Q. Fu, Y. Du, and R. H. Liu, Appl. Phys. Lett. \textbf{115}, 192403 (2019).
\bibitem{STreview}N. Locatelli, V. Cros, and J. Grollier, Nat. Mater. \textbf{13}, 11 (2014).
\bibitem{am}A. Kurenkov, S. DuttaGupta, C. L. Zhang, S. Fukami, Y. Horio, and H. Ohno, Adv. Mater. \textbf{31}, (2019).
\bibitem{switching}P. M. Haney, H.-W. Lee, K.-J. Lee, A. Manchon, and M. D. Stiles, Phys. Rev. B \textbf{87}, 174411 (2013).
\bibitem{switching2}I. M. Miron, K. Garello, G. Gaudin, P.-J. Zermatten, M. V. Costache, S. Auffret, S. Bandiera, B. Rodmacq, A. Schuhl, and P. Gambardella, Nature \textbf{476}, 189 (2011).
\bibitem{SOTintro}K. Garello, C. O. Avci, I. M. Miron, M. Baumgartner, A. Ghosh, S. Auffret, O. Boulle, G. Gaudin, and P. Gambardella, Appl. Phys. Lett. \textbf{105}, 212402 (2014).
\bibitem{SOTintro2}C. Zhang, S. Fukami, H. Sato, F. Matsukura, and H. Ohno, Appl. Phys. Lett. \textbf{107}, 012401 (2015).
\bibitem{SOTsynapse}W. A. Borders, H. Akima, S. Fukami, S. Moriya, S. Kurihara, Y. Horio, S. Sato, and H. Ohno, Appl. Phys. Express \textbf{10}, 013007 (2017).

\bibitem{SOTsynapse2}S. Fukami, and H. Ohno, J. Appl. Phys. \textbf{124}, 151904 (2018).
\bibitem{SOTsynapse3}W. A. Borders, S. Fukami, and H. Ohno, Jpn. J. Appl. Phys. \textbf{57}, 1002B2 (2018).
\bibitem{Ic}K.-S. Lee, S.-W. Lee, B.-C. Min, and K.-J. Lee, Appl. Phys. Lett. \textbf{102}, 112410 (2013).
\bibitem{Ic2}D. Zhu, and W. Zhao, Phys. Rev. Appl. \textbf{13}, 044078 (2020).
\bibitem{MAic}Y. Zhuo, W. Cai, D. Zhu, H. Zhang, A. Du, K. Cao, J. Yin, Y. Huang, K. Shi, and W. Zhao, Sci. China Phys. Mech. Astron. \textbf{65}, 107511 (2022).
\bibitem{mumax3}A. Vansteenkiste, J. Leliaert, M. Dvornik, M. Helsen, F. Garcia-Sanchez, and B. Van Waeyenberge, AIP Advances \textbf{4}, 107133 (2014).
\bibitem{SOTLLG}M. Hayashi, J. Kim, M. Yamanouchi, and H. Ohno, Phys. Rev. B \textbf{89}, 144425 (2014).
\bibitem{ST}A. Manchon, and S. Zhang, Phys. Rev. B \textbf{78}, 212405 (2008).
\bibitem{thermalfield2}E. B. Myers, F. J. Albert, J. C. Sankey, E. Bonet, R. A. Buhrman, and D. C. Ralph, Phys. Rev. Lett. \textbf{89}, 196801 (2002).
\bibitem{thermalfield3}W. F. Brown, Phys. Rev. \textbf{130}, 1677 (1963).
\bibitem{thermaltimestep}J. Leliaert, J. Mulkers, J. De Clercq, A. Coene, M. Dvornik, and B. Van Waeyenberge, AIP Advances \textbf{7}, 125010 (2017).
\bibitem{thermal}K.-S. Lee, S.-W. Lee, B.-C. Min, and K.-J. Lee, Appl. Phys. Lett. \textbf{104}, 072413 (2014).
\bibitem{probability}T. Taniguchi, and H. Imamura, Phys. Rev. B \textbf{83}, 054432 (2011).
\bibitem{COMSOL1}J. Kimling, A. Philippi-Kobs, J. Jacobsohn, H. P. Oepen, and D. G. Cahill, Phys. Rev. B \textbf{95}, 184305 (2017).
\bibitem{LIF}A. N. Burkitt, Biol. Cybern. \textbf{95}, 1 (2006).
\bibitem{STDPrule}H. Shouval, Front. Comput. Neurosc. \textbf{4}, 00019 (2010).
\bibitem{MAstdp}Q. Liu, X. Zhang, Q. Luo, X. Zhao, H. Lv, S. Long, and M. Liu, Sci. China Phys. Mech. Astron. \textbf{61}, 088711 (2018).
\bibitem{brian2}M. Stimberg, R. Brette, Dan FM Goodman, eLife \textbf{8}, e47314 (2019)
\bibitem{SNNprogram}P. U. Diehl, and M. Cook, Front. Comput. Neurosc. \textbf{9}, 00099 (2015).
\bibitem{MNIST}Y. Lecun, L. Bottou, Y. Bengio, and P. Haffner, Proc. IEEE \textbf{86}, 2278 (1998).
\bibitem{neuronsMemristor}Q. Duan, Z. Jing, X. Zou, Y. Wang, K. Yang, T. Zhang, S. Wu, R. Huang, and Y. Yang, Nat. Commun. \textbf{11}, 3399 (2020).
\bibitem{neuronsCMOS}K. Cameron, V. Boonsobhak, A. Murray, and D. Renshaw, IEEE Trans. Neural Netw. \textbf{16}, 1626 (2005).
\bibitem{SNNmtj3}A. Sengupta, P. Panda, P. Wijesinghe, Y. Kim, and K. Roy, Sci. Rep. \textbf{6}, 30039 (2016).
\bibitem{SNNmtj2}G. Srinivasan, A. Sengupta, and K. Roy, Sci. Rep. \textbf{6}, 29545 (2016).
\bibitem{logicsnn}L. Mo, and M. Wang, Electronics \textbf{10}, 2123 (2021).


\end{thebibliography}
\end{document}